\documentclass[10pt,twocolumn,letterpaper]{article}

\usepackage{cvpr}              

\usepackage{graphicx}
\usepackage{amsmath}
\usepackage{amssymb}
\usepackage{booktabs}
\usepackage{tabularx}

\usepackage[pagebackref,breaklinks,colorlinks]{hyperref}

\usepackage[capitalize]{cleveref}
\crefname{section}{Sec.}{Secs.}
\Crefname{section}{Section}{Sections}
\Crefname{table}{Table}{Tables}
\crefname{table}{Tab.}{Tabs.}

\def\cvprPaperID{9} 
\def\confName{CVPR}
\def\confYear{2022}

\begin{document}

\title{Semi-Supervised Super-Resolution}

\author{Ankur Singh\\
Indian Institute of Technology Kanpur\\
{\tt\small ankur22997@gmail.com}
\and
Piyush Rai\\
Indian Instiute of Technology Kanpur\\
{\tt\small piyush@cse.iitk.ac.in}
}
\maketitle

\begin{abstract}

Super-Resolution is the technique to improve the quality of a low-resolution photo by boosting its plausible resolution. The computer vision community has extensively explored the area of  Super-Resolution. However, previous Super-Resolution methods require vast amounts of data for training which becomes problematic in domains where very few low-resolution, high-resolution pairs might be available. One such area is statistical downscaling, where super-resolution is increasingly being used to obtain high-resolution climate information from low-resolution data. Acquiring high-resolution climate data is extremely expensive and challenging. To reduce the cost of generating high-resolution climate information, Super-Resolution algorithms should be able to train with a limited number of low-resolution, high-resolution pairs. This paper tries to solve the aforementioned problem by introducing a semi-supervised way to perform super-resolution that can generate sharp, high-resolution images with as few as 500 paired examples. The proposed semi-supervised technique can be used as a plug-and-play module with any supervised GAN-based Super-Resolution method to enhance its performance. We quantitatively and qualitatively analyze the performance of the proposed model and compare it with completely supervised methods as well as other unsupervised techniques. Comprehensive evaluations show the superiority of our method over other methods on different metrics. We also offer the applicability of our approach in statistical downscaling to obtain high-resolution climate images.
\end{abstract}

\section{Introduction}
\label{sec:intro}

A digital image consists of pixels, and the density of these pixels constitutes the spatial resolution of the image. The higher the resolution, the more refined the image details are. Ergo, a high-resolution picture is almost always desirable. However, a high-resolution image might not be available in every situation. The inability to obtain a high-resolution image can arise from the limitations of the camera, as capturing high-resolution images requires devices with better sensors having more pixels. Images can also get degraded during image compression and transfer.  \\

Single Image Super-Resolution (SISR) is the task of reconstructing high-resolution images from low-resolution ones. It is an ill-posed problem, as there is no unique high-resolution output for a low-resolution input. Lately, Super-Resolution (SR) has received much attention from the research communities and has been widely studied. Recently proposed CNN and GAN-based methods have solved the problem of generating high-resolution outputs to a great extent. \\ 

\begin{figure}[t!]
    \centering
    \begin{tabular}{ccc}
        \includegraphics[width=0.3\linewidth]{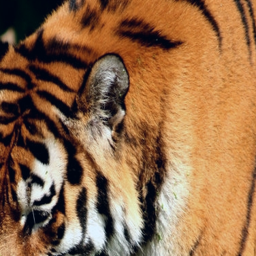}&  \includegraphics[width=0.3\linewidth]{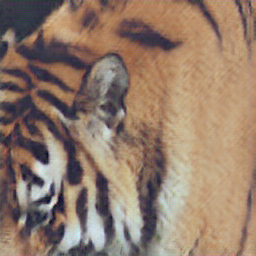}&
        \includegraphics[width=0.3\linewidth]{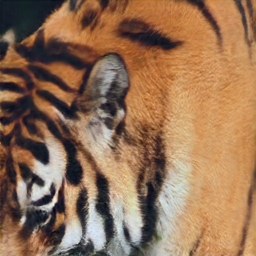} \\
         {\small (a) HR} &  {\small(b) ESRGAN} &  {\small(c) Ours}
        \\
    \end{tabular}
    \caption{Super Resolution output of ESRGAN \cite{DBLP:conf/eccv/WangYWGLDQL18} and our method when trained on 500 LR-HR image pairs
    }
    \label{fig:intro}
\end{figure}

Although deep learning-based super-resolution techniques have outperformed other non learning-based methods, the one challenge they face is the need for large amounts of training data with paired low-resolution and high-resolution images. In practice, it is challenging to obtain paired LR-HR images. Capturing HR images requires costly equipments as well as ideal conditions to shoot, and thus it becomes difficult to acquire them. The unavailability of LR-HR image pairs makes supervised learning impractical. However, with a large amount of visual data being uploaded online every day, there is no shortage of LR images. The lack of HR images and the abundance of LR images call for unsupervised techniques for super-resolution. Unfortunately, to our knowledge, not much work has been done to solve this problem in super-resolution. In this work, we tackle this problem by introducing a novel semi-supervised solution. We devise an algorithm that is able to generate super resolved images from low-resolution images using as few as 500 labeled examples. \\

Although our method can be used with any supervised SR technique, however, in this work, we revisit ESRGAN\cite{DBLP:conf/eccv/WangYWGLDQL18}, a Super-Resolution technique based on Generative Adversarial Networks, and add our unsupervised component to it. The supervised part of our method makes use of ESRGAN\cite{DBLP:conf/eccv/WangYWGLDQL18}, and the unsupervised feature utilizes consistency loss. The consistency loss can be used to generate high-resolution images from low-resolution ones without labeled pairs. Furthermore, the consistency loss enforces our belief that the generated high-resolution image, when converted to a low-resolution image, should correspond to the original low-resolution image we began with. \\

Formally the generator $G$ translates a low-resolution image $I^{LR}$ to a high-resolution image $I^{SR}$ through $G: {LR}$ $\rightarrow$ ${SR}$. $I^{SR}$ when downsampled through a downsampling function $F: {SR}$ $\rightarrow$ ${LR}$, should give back the original low-resolution image $I^{LR}$. Consistency loss for unpaired image generation has previously been used in \cite{DBLP:conf/iccv/ZhuPIE17}. However, the cycle-consistency loss in \cite{DBLP:conf/iccv/ZhuPIE17} employs two GANs. Optimizing two generators and two discriminators simultaneously can be challenging. The training process in such case also becomes memory and compute-intensive. On the other hand, our method can generate sharper images by using a single GAN thus also resulting in faster convergence.  \\

We make the following contributions in this paper:

\begin{itemize}
    \item We introduce a novel semi-supervised technique for super-resolution that can
generate high-resolution images with extremely few labeled examples.

    \item The consistency loss that we take advantage of in this work can be utilized with the current GAN-based SR models as a plug-and-play module without any change in the architecture.
    
    \item The proposed technique consists of only a single generator and one discriminator compared to previous methods that use multiple generators and discriminators to accomplish the task. 

\end{itemize}
 
\section{Related Work}
In this section, we present previous related works on the topic of Super-Resolution in
subsection \ref{rw_sr}, on Semi-Supervised Learning in subsection \ref{rw_ssl}, and on Unsupervised Super-Resolution in subsection \ref{rw_usr}

\subsection{Super-Resolution}\label{rw_sr}
This section focuses on previous deep learning-based techniques to tackle the problem of Super-Resolution. The work Image Super-Resolution Using Deep Convolutional Networks (SRCNN) by Dong et al. \cite{DBLP:conf/eccv/DongLHT14} was one of the pioneers in this area and could map LR images to HR images in an end-to-end fashion. SRCNN was the earliest works that used deep learning for SR and showed remarkable improvements in performance from its non-deep learning counterparts, thus setting a pathway for more research to follow in this domain. Inspirited by the progress of deep VGG \cite{simonyan2014very} networks, Kim et al. \cite{kim2016accurate} proposed a Very Deep Super-Resolution network (VDSR) that could learn residual images. ESPCN \cite{shi2016real} and FRCNN \cite{dong2016accelerating} sped up SR by extracting features from low-resolution photos and then upscaling them in the final layer using sub-pixel and transposed convolution, respectively. \\

The field has also observed different deep learning architectures being successfully applied. These include residual learning networks \cite{kim2016accurate}, deep laplacian pyramid structures \cite{lai2017deep}, back-projection networks\cite{haris2018deep}, recursive learning \cite{kim2016deeply}, etc. Notably, Lim et al. \cite{lim2017enhanced} presented the EDSR network that removed Batch Normalization layers. The memory saved from the removal of Batch Normalization layers was used to create larger residual blocks. As models have become deeper, new approaches have been devised to stabilize their training. Residual connection \cite{he2016deep} is one such approach that improves the performance of deep networks. Inspired by this, Wang et al. \cite{DBLP:conf/eccv/WangYWGLDQL18} use a residual-in-residual dense block to train their deep networks for SR. \\

\begin{figure*}[t!]
    \centering
     \includegraphics[height=5.5cm,width=\linewidth]{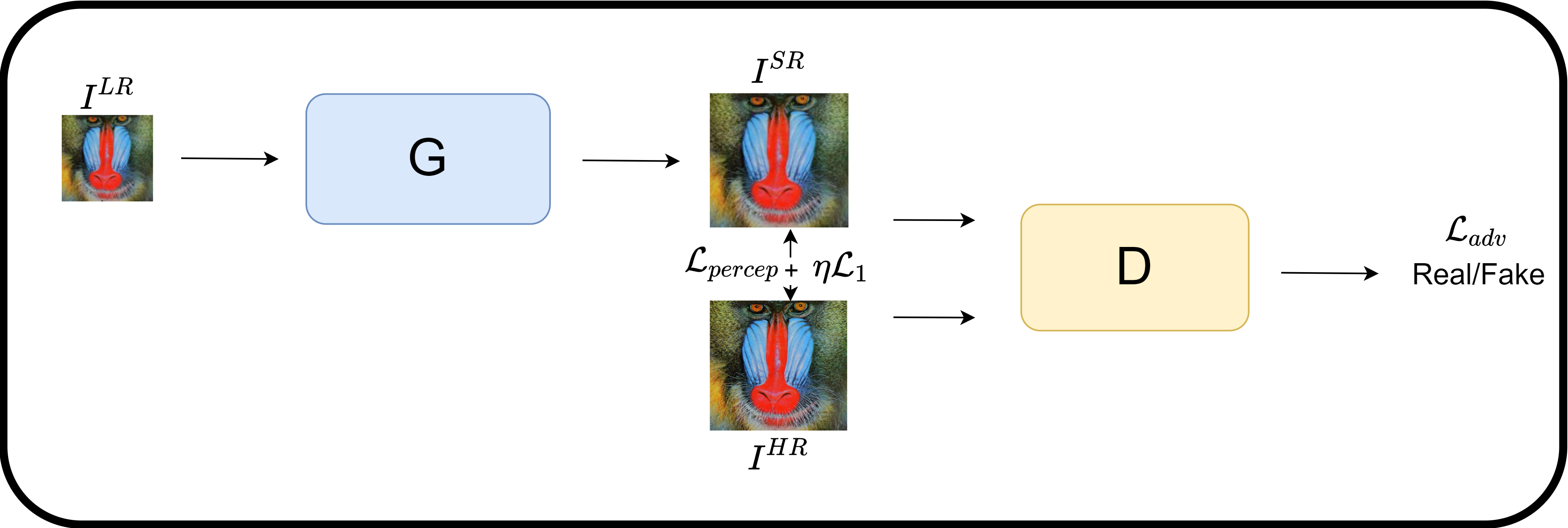}

    \caption{ESRGAN Model: LR images are passed into $G$ to generate HR images $I^{SR}$. The generated images are compared with the real HR images $I^{HR}$ using $\mathcal{L}_{percep}$ and $\mathcal{L}_1$. Along with that, $I^{SR}$ and $I^{HR}$ are also fed to $D$ to calculate $\mathcal{L}_{adv}$}
    \label{fig:esrgan}
    
\end{figure*}

\subsection{Semi-Supervised Learning}\label{rw_ssl}
 The most prevalent technique for training neural networks in a semi-supervised way is bootstrapping the model with added labeled examples generated from its own predictions. Labels obtained in this way are known as pseudo labels. Once the pseudo labels are generated, the network can be trained on the labeled and the unlabeled examples concurrently. \cite{lee2013pseudo} was the first work to adopt pseudo labels for semi-supervised learning. Methods like \cite{ranzato2008semi}, \cite{weston2012deep}, \cite{zhao2015stacked} employ auto-encoders to the network to obtain effective representations. \\

The current state of the art for image classification using semi-supervised learning are methods based on consistency regularization. A general observation is to minimize the cross-entropy loss along with the regularization loss, where the regularization loss regulates the consistency of perturbed unlabeled data. Sajadi et al. \cite{NIPS2016_30ef30b6} achieve competitive results on image classification datasets using this technique. Laine et al. \cite{laine2016temporal} propose a temporal ensembling model that attains regularization by predicting the same label output when run twice on a data point. Mean Teacher \cite{tarvainen2017mean} uses a teacher-student model where the teacher is a moving average of the training model, i.e., student. The teacher model then enforces consistency on the predictions of the student. Lately, fixmatch \cite{sohn2020fixmatch} and mixmatch \cite{berthelot2019mixmatch} introduce robust augmentations and impose consistency over them. 

\subsection{Unsupervised Super-Resolution}\label{rw_usr}
HR-LR image pairs may not always be available since capturing HR images requires expensive cameras and ideal shooting conditions. However, LR images are abundant in number, and thus it is essential to find a way to realize learning techniques that make use of these unpaired LR images. Recently GAN based methods are being heavily adopted for unsupervised image to image translation problems. CycleGAN \cite{DBLP:conf/iccv/ZhuPIE17} and DualGAN \cite{yi2017dualgan} are two such pioneer works that use unsupervised learning for image translation. Both of them propose a forward-backward generator-discriminator pair. The forward generator translates data from domain X to domain Y, whereas the backward generator translates domain Y back to domain X to enforce cycle consistency. WESPE \cite{ignatov2018wespe} uses a similar technique for weakly supervised image enhancement. \\

\begin{figure*}[t!]
    \centering
     \includegraphics[height=5.5cm,width=\linewidth]{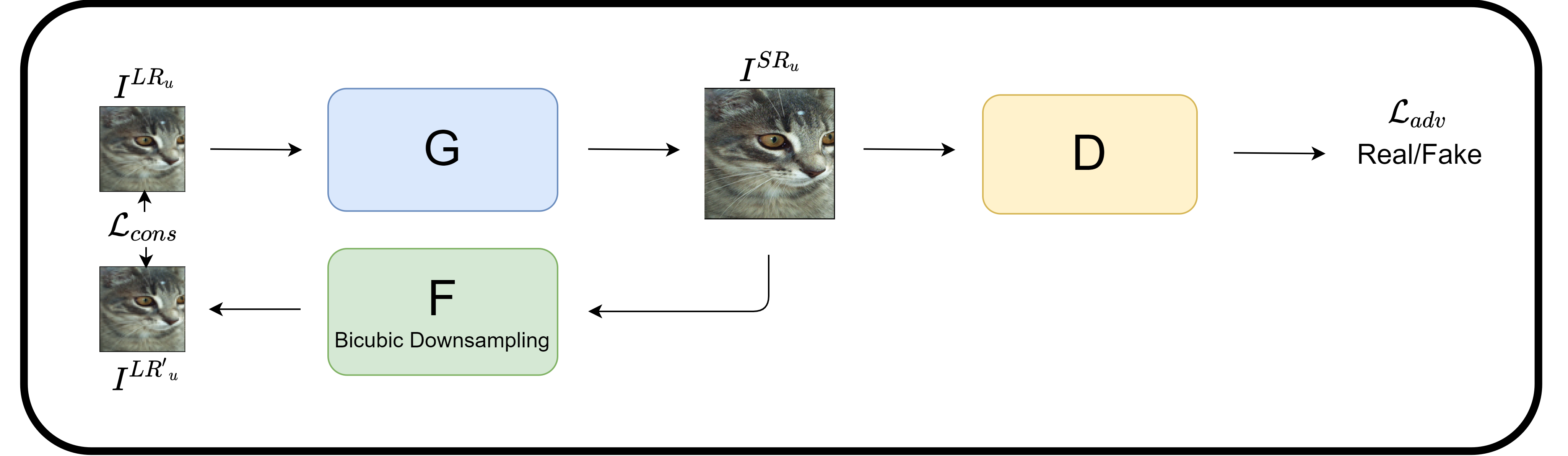}

    \caption{Unsupervised component of the model: Unpaired LR images $I^{{LR}_u}$ are passed into $G$ to generate HR images $I^{{SR}_u}$. $I^{{SR}_u}$ is then sent into $D$ to calculate $\mathcal{L}_{adv}$. Along with that, $I^{{SR}_u}$ is also downsampled using F (bicubic downsampling) to output $I^{{LR'}_u}$. The consistency loss $\mathcal{L}_{cons}$ is then computed between $I^{{LR}_u}$ and $I^{{LR'}_u}$.} 
    \label{fig:esrgan_ssl}
    
\end{figure*}

Image to image translation methods generate output images that are of the same size as their inputs. However, by definition, SR outputs are several factors larger than their inputs, and it is therefore challenging to perform SR with paired images, let alone with unpaired ones. Cycle-in-Cycle GAN (CinCGAN) \cite{yuan2018unsupervised} is one work that focuses on accomplishing unsupervised SR by implementing a network consisting of three generators and two discriminators. Recently proposed \cite{hou2022semi} makes use of a pre-trained SR network and the CycleGAN architecture consisting of two generators and two discriminators for super-resolution, making it exceptionally computationally expensive to train. On the other hand, the proposed architecture in this paper consists of only one generator and one discriminator, which drastically reduces the number of parameters and is consequently easier to train. Utilizing the characteristics of cycle consistency loss, we offer a semi-supervised way to perform SR. Our method can be used as a plug-and-play module with the current supervised SR architectures and can generate sharp, realistic, and high-frequency HR images.

\section{Proposed Approach}
Single Image Super-Resolution (SISR) aims to generate high-quality SR images $I^{SR}$ from low-resolution images $I^{LR}$. The problem formulation in SISR can be written as: 
\begin{equation}
 I^{LR} = SHI^{HR} + n
\end{equation}  

Here $I^{LR}$ denotes LR images, and $I^{HR}$ denotes HR images. $S$ and $H$ signify the down-sampling and the blurring matrix, respectively, and $n$ represents the noise. \\

In the case of supervised SR, high-resolution ground truth images, $I^{HR}$ are also provided. However, we have only $L$ $I^{LR}-I^{HR}$ image pairs and $U$ unpaired $I^{LR}$ images for semi-supervised SR. For this problem, let $G$ be the generator, and $\theta_{G}$ be its parameters, then our goal is to generate $I^{SR}$ from $G$ through $G(I^{LR}, \theta_{G})$. The generator should be able to carry out blind SR and reconstruct visually pleasing $I^{SR}$ images similar to HR ground truths. Let $\mathcal{L}_{s}$ be the supervised component and $\mathcal{L}_{u}$ be unsupervised part of the loss function, then we have the following optimization problem: 

\begin{equation}
\begin{aligned}
  & \min_{\theta_G} (\sum_{(I^{LR},I^{HR}) \epsilon X_{L}}\mathcal{L}_{s}(G(I^{LR}; \theta_{G}), I^{HR})  + \\
  & \sum_{(I^{LR}) \epsilon X_{U}}\alpha\mathcal{L}_{u}(G(I^{LR}; \theta_{G})))
  \end{aligned}
\end{equation}

Our main objective is to devise a semi-supervised technique to solve the problem of SR. For this reason, we build upon the work done in ESRGAN \cite{DBLP:conf/eccv/WangYWGLDQL18} and add a semi-supervised component to it. Although in this work, we focus on making ESRGAN semi-supervised, however, our method can be applied to any other GAN-based SR architecture as well. \\

ESRGAN uses an adversarial network architecture for SR. The idea is to generate SR images through a generator $G$ that can fool a discriminator $D$ trained for differentiating real HR images from generated HR images. Once trained, the generator is able to create HR images that look sharp and realistic, and thus it becomes challenging for the discriminator to classify them correctly. The adversarial network model encourages perceptually pleasing images, unlike the results obtained by minimizing pixel-based loss functions that were predominant earlier. \\

The framework of the ESRGAN model is shown in Figure: \ref{fig:esrgan}. The model consists of a generator $G$ and a discriminator $D$. LR images are first fed to the generator $G$, which then outputs HR images $I^{SR}$. The error between the generated HR images $I^{SR}$ and the ground truth HR images $I^{HR}$ is then calculated using the perceptual loss $\mathcal{L}_{{percep}_s}$ and the pixel-wise L1 loss $\mathcal{L}_{{1}_s}$. Apart from this, $I^{SR}$ and $I^{HR}$ are also passed through the discriminator $D$ to compute the adversarial loss $\mathcal{L}_{{adv}}$. The discriminative component $\mathcal{L}_{{adv}_D}$ of $\mathcal{L}_{{adv}}$ is used to update the parameters of $D$ while the parameters of $G$ get updated by utilizing $\mathcal{L}_{{percep}_s}$, $\mathcal{L}_{{1}_s}$ and the generative component of $\mathcal{L}_{{adv}}$ that is $\mathcal{L}_{{adv}_{{G}_s}}$. Here the subscript $s$ represents loss functions in supervised training where paired images are available. \\

The total loss of the generator $\mathcal{L}_{{G}_s}$ is a combination of the three losses mentioned above. The resulting loss function  is:
\begin{equation}
    \mathcal{L}_{{G}_s} = \mathcal{L}_{{percep}_s} + \lambda\mathcal{L}_{adv_{{G}_s}} + \eta
    \mathcal{L}_{{1}_s}
\end{equation}

Here $\lambda$ and $\eta$ are the weights for $\mathcal{L}_{adv_{{G}_s}}$ and $\mathcal{L}_{{1}_s}$ respectively. \\

We now describe our novel semi-supervised technique that can be added to any GAN-based SR architecture. However, in this work, we mainly focus on utilizing it along with ESRGAN. The framework of the SSL component is shown in Figure: \ref{fig:esrgan_ssl}. The SSL method uses a combination of adversarial loss $\mathcal{L}_{{adv}}$ and consistency loss $\mathcal{L}_{cons}$ described below: 

\textbf{Adversarial loss}
In our SSL method, $I^{{LR}_u}$ is also passed to the generator $G$ along with $I^{{LR}}$ to generate $I^{{SR}_u}$ and $I^{{SR}}$ as the outputs. $I^{{SR}_u}$, $I^{{SR}}$ and $I^{{HR}}$ are then sent to the discriminator to classify them as real or fake. The complete adversarial loss for the generator $\mathcal{L}_{{adv}_G}$ now consists of $\mathcal{L}_{{adv}_{Gs}}$ and $\mathcal{L}_{{adv}_{Gu}}$. Here the subscript $u$ denotes that the loss function has been used for unsupervised training. 
\begin{equation}
    \mathcal{L}_{{adv}_G} = \mathcal{L}_{{adv}_{Gs}} + \mathcal{L}_{{adv}_{Gu}}
\end{equation}

Here $\mathcal{L}_{{adv}_{Gs}} = -\log D(G(I^{LR}))$ and $\mathcal{L}_{{adv}_{Gu}} = -\log D(G(I^{LR_{u}}))$ \\

Adversarial learning can enforce the generator to map unpaired images from the LR domain to the SR domain. However, with the large enough capacity, the unpaired images can be mapped to any random perturbation of the target domain. To ensure that $I^{{LR}_u}$ gets mapped to its own high-resolution variant $I^{{SR}_u}$ and not to any other perturbation in the target domain, we also introduce a consistency loss. \\

\textbf{Consistency Loss:} The consistency loss brings a way of supervision for the unpaired images to reduce the chance of inducing undesirable variations in the reconstructed output. We argue that the reconstructed SR output $I^{{SR}_u}$ when mapped back to the LR domain should be consistent with $I^{{LR}_u}$ that we started with. Formally, $I^{{SR}_u}$ when downsampled through a downsampling function $F: SR$ $\rightarrow$ $LR$, should give back the original low-resolution image $I^{{LR}_u}$. Mathematically, this can be formulated as: 
\begin{equation}
    I^{{LR}_u} \rightarrow G \rightarrow G(I^{{LR}_u}) \rightarrow F \rightarrow F(G(I^{{LR}_u})) \approx I^{{LR}_u}
\end{equation}

We implement the consistency loss $\mathcal{L}_{{cons}}$ as a combination of pixel-wise L1 loss $\mathcal{L}_{{1}_u}$ and perceptual loss $\mathcal{L}_{{percep}_u}$. 
\begin{equation}
    \mathcal{L}_{{cons}} =   \alpha\mathcal{L}_{{percep}_u} + \beta\mathcal{L}_{{1}_u}
\end{equation}

Where $\mathcal{L}_{{1}_u}$ can be calculated as:
\begin{equation}
     \mathcal{L}_{{1}_u} = {\dfrac{1}{W_{LR}H_{LR}}}\sum_{{x=1}}^{W_{LR}}\sum_{{y=1}}^{H_{LR}}\left\lVert{I^{{LR}_u}_{(x,y)}}-F(G(I^{{LR}_u}))_{(x,y)}\right\rVert
\end{equation} 
Here, $W_{LR}$ and $H_{LR}$ are the width and the height of the LR image, respectively. \\

And $\mathcal{L}_{{percep}_u}$ can be calculated using the following mathematical formula: 
\begin{equation}
\begin{aligned}
  & \mathcal{L}_{{percep}_u} = \\
    \hspace{-1.75em}& {\dfrac{1}{W_{i,j}H_{i,j}}}\sum_{{x=1}}^{W_{i,j}}\sum_{{y=1}}^{H_{i,j}}\left\lVert{\phi_{i,j}(I^{{LR}_u})_{(x,y)}}-\phi_{i,j}(F(G(I^{{LR}_u})))_{(x,y)}\right\rVert
\end{aligned}
\end{equation}
Here $\phi_{i,j}$ represents the features obtained from the $j^{th}$ convolution before the $i^{th}$ max-pooling layer. $W_{i,j}$ and $H_{i,j}$ denote the dimensions of the feature maps. For the particular case of ESRGAN, $j$ is 4, and $i$ is 5. \\

\textbf{Final Objective function - Proposed method}
In the proposed method, the total loss of the generator $\mathcal{L}_{{G}}$ is a sum of the supervised loss $\mathcal{L}_{{G}_s}$ and the unsupervised loss $\mathcal{L}_{{G}_u}$.

Which can be written as:
\begin{equation} \label{obj_func}
\begin{aligned}
      \hspace{-3em}&\mathcal{L}_{{G}} = \mathcal{L}_{{percep}_s} + \lambda\mathcal{L}_{adv_{{G}_s}} + \\
      & \eta\mathcal{L}_{{1}_s} + \alpha\mathcal{L}_{{percep}_u} + \gamma\mathcal{L}_{adv_{{G}_u}} + \beta
    \mathcal{L}_{{1}_u}
\end{aligned}
\end{equation} 

\section{Experiments and Results}
In this section, we describe the experiments and the corresponding results.
\subsection{Datasets}
We perform experiments on the Outdoor Scenes Train/Test (OST) \cite{wang2018sftgan} Animals dataset and the 2D surface temperature forecasts dataset obtained from Environment and Climate Change Canada (ECCC) \footnote{https://www.canada.ca/en/environment-climate-change.html}. \\

\textbf{OST Dataset:} The OST Animals dataset consists of 2187 high-resolution images of animals. The images
present in the dataset are of varying sizes. Since the Animals dataset doesn't have a test
set, we shuffle the dataset and use 238 images for testing and 1949 images for training. \\

\textbf{ECCC surface temperature dataset:}
The ECCC 2D surface temperature dataset is a statistical downscaling dataset. Statistical downscaling is a task where the objective is to obtain climate information at large scales and use it to predict information at local scales. The 2D surface temperature dataset contains 2D surface temperature forecasts obtained from ECCC's weather forecast models. The dataset consists of 5343 training HR-LR image pairs and 248 test images. The scale factor between the HR-LR pairs is 4 (10 km for LR images and 2.5 km for HR images). The size of both LR as well as HR images is 256 x 256. \\

\subsection{Implementation Details}
We train our networks on an NVIDIA GTX 1080 GPU. We use Pytorch \cite{paszke2019pytorch} for all our experiments. For the OST dataset, the LR images are obtained through bicubic downsampling by a factor of 4. In the ECCC 2D surface temperature dataset, the LR images are provided but have a shape of 256 x 256. All the LR images are downsampled to a shape 64 x 64, while the HR images have a shape 256 x 256. \\

We initialize the learning rate to $2\times 10^{-4}$. Similar to ESRGAN, we also follow a two-stage training process to train our networks. We start by training the generator using a PSNR-based model with the L1 loss for the first 500 batches. After this stage, the generator is trained using the loss function introduced in Equation \ref{obj_func} with $\lambda = 2.5\times 10^{-3}$, $\eta = 10^{-2}$, $\alpha = 10^{-1}$, $\gamma = 2.5\times 10^{-3}$, and $\beta = 5\times 10^{-3}$. The usage of pixel-wise loss as a warmup for the 500 batches helps generate more visually pleasing outputs since after-pretraining with the L1 loss; the discriminator receives more realistic super-resolved images from the generator rather than extremely fake-looking images. \\

We use the famous algorithm, Adam\cite{kingma2014adam}, for optimization with $\beta_{1} = 0.9$ and $\beta_{2} = 0.999$. The parameters of the generator and the discriminator are updated alternately until convergence. The generator in our experiments consists of 23 RRDB blocks. The features for the perceptual loss are extracted from the 4th convolutional layer before the 5th max-pool using a pre-trained VGG-19.  \\

\subsection{Metrics} \label{metrics}
We evaluate the performance of our approach based on the following metrics:

\textbf{Frechet Inception Distance} \cite{heusel2017gans} or FID is a metric to measure the quality of the outputs generated from generative networks, such as GANs. FID compares the generated outcomes from a GAN with real images. It calculates the squared Wassertian metric between two multidimensional Gaussian distributions. The two distributions usually are the neural network features of real images and the neural network features of the generated images. A pre-trained Inception is most commonly used as the neural network to extract features from the two sets of images.
The FID score better correlates with human judgment as it is not based on pixel-based comparisons and instead compares the mean and standard deviation of the features provided by an intermediate layer of a pre-trained image classification network. \\

\textbf{Mean Opinion Score}
Mean Opinion Score (MOS) quantifies the performance of the different generative networks through human judgment. To calculate MOS, we asked 12 raters to score the outputs of the different algorithms from 1 (worst) to 5 (best). The raters were asked to score 5 versions of 10 super-resolved images: ESRGAN, CycleGAN, the proposed method, and two ablations of our method. In total, the raters rated 50 images. \\

\subsection{Training Set distribution} \label{training_set_dist}
We compare our proposed model with ESRGAN \cite{DBLP:conf/eccv/WangYWGLDQL18}, CycleGAN \cite{DBLP:conf/iccv/ZhuPIE17}, as well as two ablations of our method. In the OST dataset, we utilize 500 paired images for the supervised setting, while the unsupervised component uses 1449 images. In completely unsupervised methods, all 1949 images are unpaired, whereas completely supervised methods utilize 500 paired images (see Table \ref{table1}). Similarly, for the ECCC dataset, our proposed method uses 500 paired images and 4843 unpaired images. The completely unsupervised methods utilize 5343 unpaired images while the supervised methods use 500 paired images (see Table \ref{table2}). \\

\begin{table}[h]
\centering
\begin{tabular}{| m{5em} | m{5em}| m{5em} | m{5em} |}
\hline
\textbf{Method} & \textbf{Setting} & \textbf{Paired images} & \textbf{Unpaired images} \\ \hline
ESRGAN \cite{DBLP:conf/eccv/WangYWGLDQL18}    & Supervised         & 500                           & 0                               \\ \hline
CycleGAN \cite{DBLP:conf/iccv/ZhuPIE17}    & Unsupervised      & 0                             & 1949                            \\ \hline
Our method    & Semi-Supervised     & 500                           & 1449                            \\ \hline

\end{tabular}
\caption{Distribution of images in different settings for OST dataset}
\label{table1}
\end{table}

\begin{table}[h]
\centering
\begin{tabular}{| m{5em} | m{5em}| m{5em} | m{5em} |}
\hline
\textbf{Method} & \textbf{Setting} & \textbf{Paired images} & \textbf{Unpaired images} \\ \hline
ESRGAN \cite{DBLP:conf/eccv/WangYWGLDQL18}    & Supervised         & 500                           & 0                               \\ \hline
CycleGAN \cite{DBLP:conf/iccv/ZhuPIE17}    & Unsupervised      & 0                             & 5343                            \\ \hline
Our method    & Semi-Supervised     & 500                           & 4843                           \\ \hline

\end{tabular}
\caption{Distribution of images in different settings for ECCC dataset}
\label{table2}
\end{table}

\subsection{Quantitative Results} \label{quant}
To study the effect of the consistency loss in our method, we propose two ablations. The first ablation (Ablation 1) doesn't use the consistency loss $\mathcal{L}_{cons}$ presented in this work. The second ablation (Ablation 2) doesn't use the perceptual loss $\mathcal{L}_{{percep}_u}$ in the consistency loss. We analyze and evaluate the performance of the ablations along with other methods in Section \ref{fid} and Section \ref{mos}.

\subsubsection{FID scores} \label{fid}

\begin{table}[h]
\centering
\begin{tabular}{|l|l|}
\hline
\textbf{Algorithm} & \textbf{FID Score} 
\\ \hline
ESRGAN \cite{DBLP:conf/eccv/WangYWGLDQL18}    & 64.68                              
\\ \hline
CycleGAN \cite{DBLP:conf/iccv/ZhuPIE17}     & 111.27                            
                       
\\ \hline
Ablation 1  & 63.32                         
\\ \hline
Ablation 2  & 60.18  
\\ \hline
\textbf{Our method}  & \textbf{56.83}  
\\ \hline

\end{tabular}
\caption{FID scores of different methods on OST dataset (lower the better)}
\label{table3}
\end{table}

\begin{table}[h]
\centering
\begin{tabular}{|l|l|}
\hline
\textbf{Algorithm} & \textbf{FID Score} 
\\ \hline
ESRGAN \cite{DBLP:conf/eccv/WangYWGLDQL18}   & 23.85                              \\ \hline
CycleGAN \cite{DBLP:conf/iccv/ZhuPIE17}    & 53.77                           \\ \hline
Ablation 1  & 20.42                         
\\ \hline
Ablation 2  & 17.94                         
\\ \hline
\textbf{Our method}  & \textbf{15.37}                         
\\ \hline
\end{tabular}
\caption{FID scores of different methods on ECCC dataset (lower the better)}
\label{table4}
\end{table}

The FID scores (see Table \ref{table3} and \ref{table4}) clearly show that our method outperforms ESRGAN \cite{DBLP:conf/eccv/WangYWGLDQL18} and CycleGAN \cite{DBLP:conf/iccv/ZhuPIE17}. This proves the efficacy of our semi-supervised method that can improve any GAN-based SR technique. While ESRGAN \cite{DBLP:conf/eccv/WangYWGLDQL18} shows an FID score of 64.68 on the OST dataset, our proposed algorithm exceeds its performance and gives a score of 56.83, which is an improvement by a score of 7.85. Like ESRGAN, even our method uses 500 paired images to train in a supervised manner; however, we also utilize unpaired LR images. The objective of our proposed approach is not to outperform supervised methods but to use our technique as a semi-supervised component on top of already established supervised GAN-based SR methods and improve their results. We accomplish this objective by utilizing the same number of paired images as ESRGAN and then enhancing its performance by a significant margin.\\

The results of the ablation methods, when compared to that of ESRGAN, indicate that employing unpaired images helps improve FID scores. Even using an adversarial loss $\mathcal{L}_{adv_{{G}_u}}$ for the unpaired images gets notable performance improvements. Ablation 2 demonstrates the importance of the proposed consistency loss for our semi-supervised technique. On the other hand, the poor scores of CycleGAN \cite{DBLP:conf/iccv/ZhuPIE17} on both datasets show that completely unsupervised methods are far off from replacing their supervised counterparts for SR problems.

\subsubsection{MOS} \label{mos}

\begin{table}[h]
\centering
\begin{tabular}{|l|l|}
\hline
\textbf{Algorithm} & \textbf{MOS} 
\\ \hline
ESRGAN \cite{DBLP:conf/eccv/WangYWGLDQL18}   & 2.39                              
\\ \hline
CycleGAN \cite{DBLP:conf/iccv/ZhuPIE17}    & 1.025                           
\\ \hline

Ablation 1  & 3.43                         
\\ \hline
Ablation 2  & 3.69                         
\\ \hline
\textbf{Our method}  & \textbf{4.45}                         
\\ \hline

\end{tabular}
\caption{MOS of different methods on OST dataset (higher the better)}
\label{table5}
\end{table}

\begin{table}[h]
\centering
\begin{tabular}{|l|l|}
\hline
\textbf{Algorithm} & \textbf{MOS} 
\\ \hline
ESRGAN \cite{DBLP:conf/eccv/WangYWGLDQL18}   & 2.40                              
\\ \hline
CycleGAN \cite{DBLP:conf/iccv/ZhuPIE17}    & 1.10                         
\\ \hline

Ablation 1  & 3.10                        
\\ \hline
Ablation 2  & 3.70                        
\\ \hline
\textbf{Our method}  & \textbf{4.70}                         
\\ \hline

\end{tabular}
\caption{MOS of different methods on ECCC dataset (higher the better)}
\label{table6}
\end{table}

To calculate MOS, we used 10 sets of images, each set containing outputs from the 5 different generative models, including an HR ground truth for reference. Raters were asked to score the generated images between 5 (best) to 1 (worst). The MOS results are presented in Table \ref{table5} and \ref{table6}. The scores indicate that the human reviewers found the generated output of our model to be more realistic and similar to the HR image. Ablation 2 was the 2nd choice of most of the reviewers, closely followed by Ablation 1. The MOS results are clear evidence to prove the effectiveness of our semi-supervised approach. While ESRGAN requires a massive number of LR-HR image pairs for training, on the other hand, our technique can generate sharp and realistic images with a significantly fewer number of image pairs.

\subsection{Qualitative Results} \label{qual}
\begin{figure*}[!htbp]
    \centering
    \begin{tabular}{cccccc}
        \includegraphics[width=0.15\linewidth]{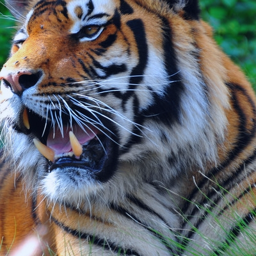}&  \includegraphics[width=0.15\linewidth]{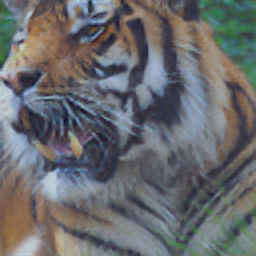}&  \includegraphics[width=0.15\linewidth]{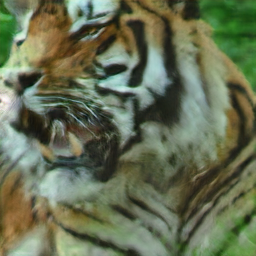}& 
        \includegraphics[width=0.15\linewidth]{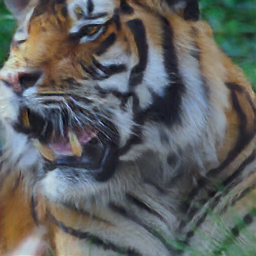}&  \includegraphics[width=0.15\linewidth]{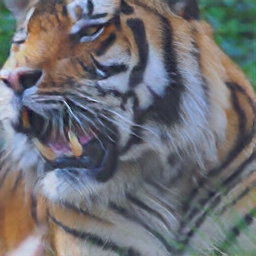}& \includegraphics[width=0.15\linewidth]{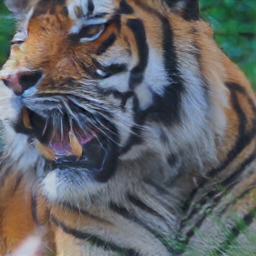} \\
         {\small (a) HR} &   {\small(b) ESRGAN} & {\small (c) CycleGAN} & {\small (d) Ablation 1} &   {\small(e) Ablation 2} & {\small (f) \textbf{Ours}}
        \\
    \end{tabular}
    \caption{
    }
    \label{fig:qual_res1}
\end{figure*}

\begin{figure*}[!htbp]
    \centering
    \begin{tabular}{cccccc}
        \includegraphics[width=0.15\linewidth]{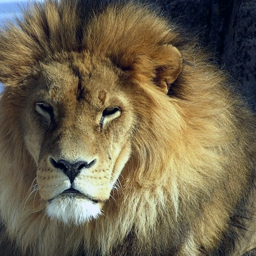}&  \includegraphics[width=0.15\linewidth]{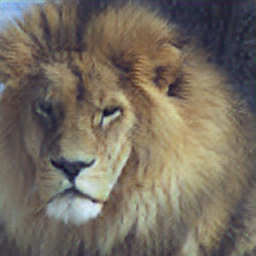}&  \includegraphics[width=0.15\linewidth]{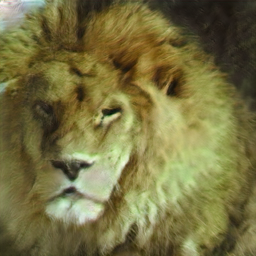}& 
        \includegraphics[width=0.15\linewidth]{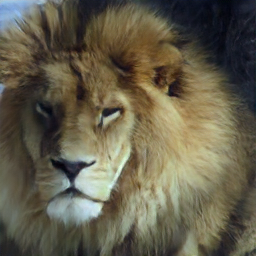}&  \includegraphics[width=0.15\linewidth]{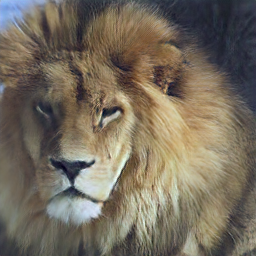}& \includegraphics[width=0.15\linewidth]{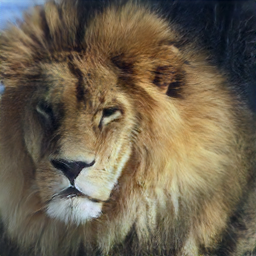} \\
         {\small (a) HR} &   {\small(b) ESRGAN} & {\small (c) CycleGAN} & {\small (d) Ablation 1} &   {\small(e) Ablation 2} & {\small (f) \textbf{Ours}}
        \\
    \end{tabular}
    \caption{
    }
    \label{fig:qual_res2}
\end{figure*}

\begin{figure*}[!htbp]
    \centering
    \begin{tabular}{cccccc}
        \includegraphics[width=0.15\linewidth]{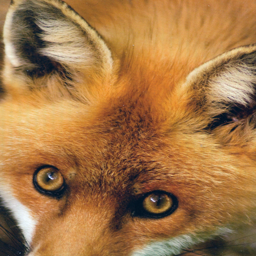}&  \includegraphics[width=0.15\linewidth]{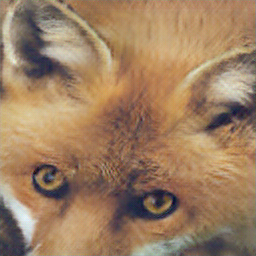}&  \includegraphics[width=0.15\linewidth]{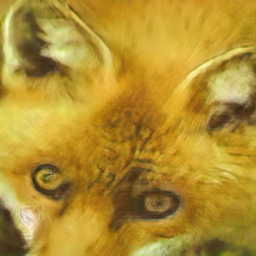}& 
        \includegraphics[width=0.15\linewidth]{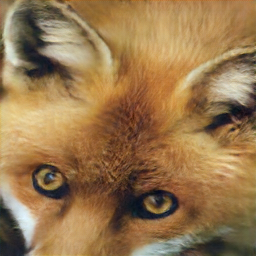}&  \includegraphics[width=0.15\linewidth]{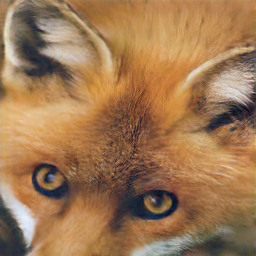}& \includegraphics[width=0.15\linewidth]{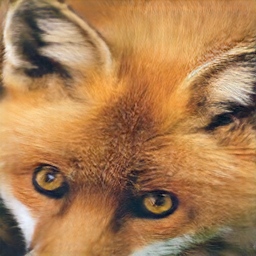}  \\
         {\small (a) HR} &   {\small(b) ESRGAN} & {\small (c) CycleGAN} & {\small (d) Ablation 1} &   {\small(e) Ablation 2} & {\small (f) \textbf{Ours}}
        \\
    \end{tabular}
    \caption{
    }
    \label{fig:qual_res3}
\end{figure*}

\begin{figure*}[!htbp]
    \centering
    \begin{tabular}{cccccc}
        \includegraphics[width=0.15\linewidth]{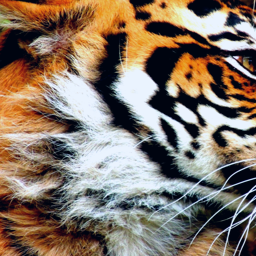}&  \includegraphics[width=0.15\linewidth]{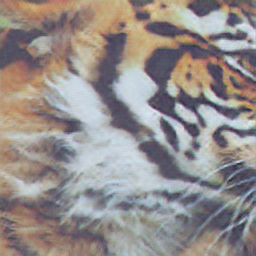}&  \includegraphics[width=0.15\linewidth]{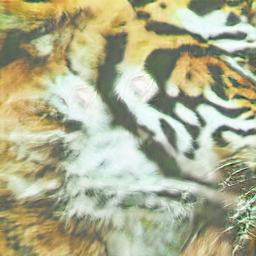}& \includegraphics[width=0.15\linewidth]{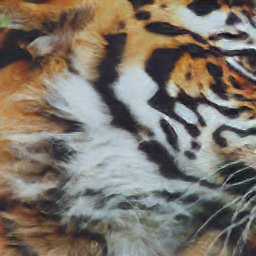}&  \includegraphics[width=0.15\linewidth]{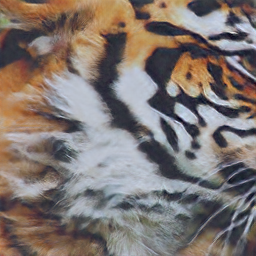}& \includegraphics[width=0.15\linewidth]{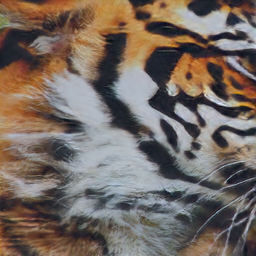}  \\
         {\small (a) HR} &   {\small(b) ESRGAN} & {\small (c) CycleGAN} & {\small (d) Ablation 1} &   {\small(e) Ablation 2} & {\small (f) \textbf{Ours}}
        \\
    \end{tabular}
    \caption{
    }
    \label{fig:qual_res4}
\end{figure*}

\begin{figure*}[!htbp]
    \centering
    \begin{tabular}{cccccc}
        \includegraphics[width=0.15\linewidth]{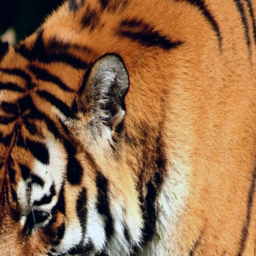}&  \includegraphics[width=0.15\linewidth]{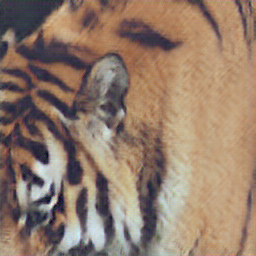}&  \includegraphics[width=0.15\linewidth]{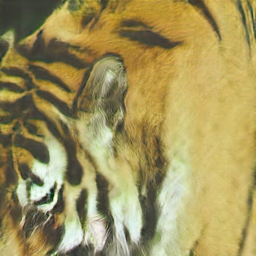}& \includegraphics[width=0.15\linewidth]{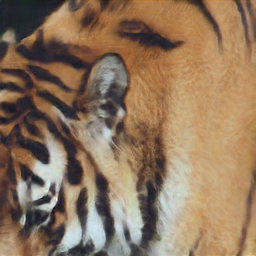}&  \includegraphics[width=0.15\linewidth]{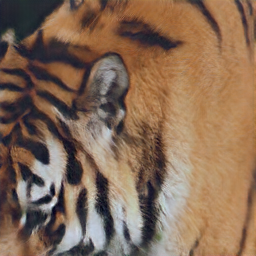}& \includegraphics[width=0.15\linewidth]{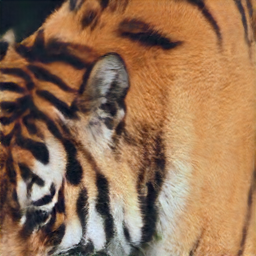}  \\
         {\small (a) HR} &   {\small(b) ESRGAN} & {\small (c) CycleGAN} & {\small (d) Ablation 1} &   {\small(e) Ablation 2} & {\small (f) \textbf{Ours}}
        \\
    \end{tabular}
    \caption{ Qualitative Results on OST dataset
    }
    \label{fig:qual_res5}
\end{figure*}

\begin{figure*}[!htbp]
    \centering
    \begin{tabular}{ccc}
        \includegraphics[width=0.3\linewidth]{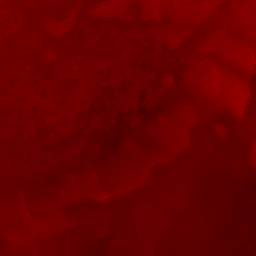}&  \includegraphics[width=0.3\linewidth]{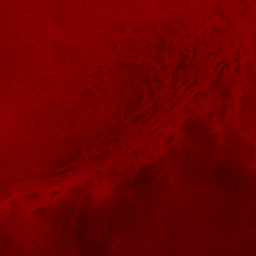}&  \includegraphics[width=0.3\linewidth]{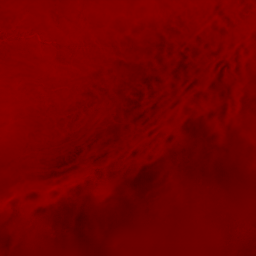} \\
        {\small (a) LR} &   {\small(b) HR} & {\small (c) \textbf{Ours}}\\
    \end{tabular}
    \caption{
    }
    \label{fig:qual_res7}
\end{figure*}

In this section, we present some qualitative results and compare our model with other approaches. We provide outputs of different generative models in figures \ref{fig:qual_res1} - \ref{fig:qual_res7}, and discuss them in some detail below. \\

Figure \ref{fig:qual_res1} - \ref{fig:qual_res5} provide the outputs of different models on the images from OST dataset. Figure \ref{fig:qual_res1} (f) - \ref{fig:qual_res5} (f) show the rich texture that the output of our model possesses. Outputs generated by our model can retain the true colors, are sharper, and have more high-frequency details than other methods. While the results of ESRGAN are pale and blurry, the Ablation 1 outputs, although not sharp, do show bright colors. This indicates that even an adversarial loss for the unpaired images is helpful and can result in notable performance improvements. Ablation 2 outputs contain high-frequency details but, in some cases, may have faded colors. The effectiveness of the proposed consistency loss can be witnessed in the results of our model, which look very similar to the original HR output. On the other hand, outputs of CycleGAN are blurry, faded, and have undesirable artifacts, which proves that current unsupervised methods are not applicable for SR problems. \\

\begin{figure*}[!htbp]
    \centering
    \begin{tabular}{ccc}
\includegraphics[width=0.3\linewidth]{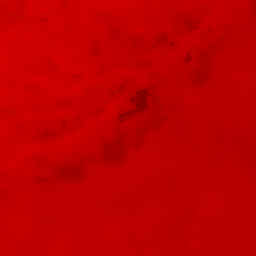}&  
\includegraphics[width=0.3\linewidth]{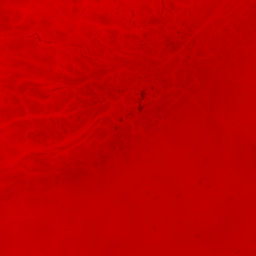}&  
\includegraphics[width=0.3\linewidth]{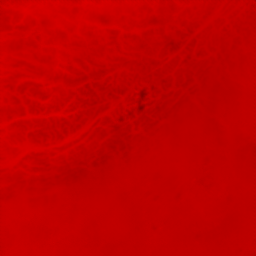} \\
{\small (a) LR} &   {\small(b) HR} & {\small (c) \textbf{Ours}} \\
    \end{tabular}
    \caption{ Qualitative Result on ECCC dataset
    }
    \label{fig:qual_res8}
\end{figure*}

\begin{figure*}[!htbp]
    \centering
    \begin{tabular}{ccc}
\includegraphics[width=0.3\linewidth]{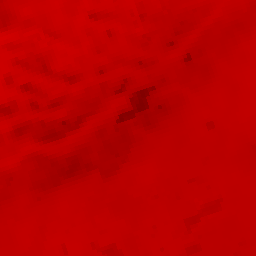}&  
\includegraphics[width=0.3\linewidth]{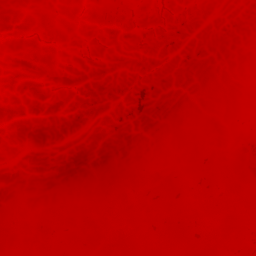}&  
\includegraphics[width=0.3\linewidth]{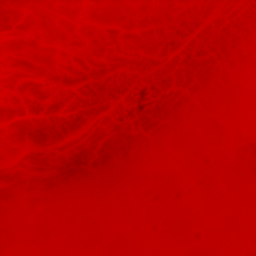} \\
{\small (a) LR} &   {\small(b) HR} & {\small (c) \textbf{Ours}} \\
    \end{tabular}
    \caption{ Qualitative Result on ECCC dataset
    }
    \label{fig:qual_res9}
\end{figure*}

In the ECCC dataset, the large-scale information has been captured at 10km, and the aim is to predict information at 2.5 km height, which is a downscaling factor of 4. HR climate images are difficult to obtain; thus, it makes our semi-supervised method that can be trained with much fewer images even more helpful in the field of statistical downscaling. Outputs generated from our model (Figure \ref{fig:qual_res7} - \ref{fig:qual_res9}) look strikingly similar to the HR output as even compared by our quantitative results hence paving a new way in the direction of semi-supervised statistical downscaling. \\

\section{Conclusion and Future Work}
In this paper, we investigated the task of Super-Resolution in a semi-supervised way. We propose a novel semi-supervised technique to tackle a more general assumption in SR where LR-HR image pairs aren't readily available. For this purpose, we present a consistency loss that can be utilized with the current model as is, without any change in the network architecture. We show that our proposed approach can be added as a semi-supervised component to any GAN-based SR method to enhance its performance significantly. Our semi-supervised method uses as few as 500 paired examples and still manages to generate outputs with bright colors, rich textures, and high-frequency features. Later, we compare our method with other baselines and ablations on various metrics to prove the effectiveness of our approach quantitatively. We also examine the outputs of different generative models in some detail and perform a qualitative analysis of them. Finally, we demonstrate the application of our semi-supervised method for Statistical Downscaling (SD) to generate high-resolution climate images that are indistinguishable from the actual HR images. We hope that our work paves a new direction to solve and overcome the requirement of a massive number of paired data for SR as well as SD. 

Our model has been tuned for GAN-based SR methods; however, in the future, it would be interesting to see if it can be extended to other SR techniques. Another exciting extension would be to make it completely unsupervised and still obtain comparable performances. Finally, much of our focus will be on entirely solving the problem of SD in an unsupervised way, where getting HR climate images is a major issue.

{\small
\bibliographystyle{unsrt}
\bibliography{egbib}
}

\end{document}